\documentclass[aps,pra,english,twocolumn,superscriptaddress]{revtex4-1}
\usepackage{graphicx}
\usepackage{amssymb}
\usepackage{amsmath, amstext, amssymb, amsfonts, amsxtra, mathtools, stmaryrd} 
\usepackage{braket}
\usepackage{textcomp}
\usepackage{relsize}
\usepackage[usenames,dvipsnames]{color}

\usepackage{units}
\usepackage[normalem]{ulem}

\newcommand{\bonn}{Physikalisches Institut, University of Bonn, Nussallee 12, 53115 Bonn, Germany}
\newcommand{\cologne}{Institute for Theoretical Physics, University of Cologne, 50937 Cologne, Germany}

\let\oldref\ref
\renewcommand{\ref}[1]{\oldref{#1}}

\begin{document}
\title{Non-equilibrium metastable state in a chain of interacting spinless fermions with localized loss}

\author{Stefan Wolff}
\affiliation{\bonn}
\author{Ameneh Sheikhan}
\affiliation{\bonn}
\author{Sebastian Diehl}
\affiliation{\cologne}
\author{Corinna Kollath}
\affiliation{\bonn}

\begin{abstract}
We investigate a chain of spinless fermions with nearest-neighbour interactions that are subject to a local loss process. We determine the time evolution of the system using matrix product state methods. We find that at intermediate times a metastable state is formed, which has very different properties than usual equilibrium states. In particular, in a region around the loss, the filling is reduced, while Friedel oscillations with a period corresponding to the original filling continue to exist. The associated momentum distribution is emptied at all momenta by the loss process and the Fermi edge remains approximately at its original value. Even in the presence of strong interactions, where a redistribution by the scattering is naively expected, such a regime can exist over a  long time-scale. Additionally, we point out the existence of the interaction dependent Zeno effect in such a system. 

\end{abstract}

\maketitle
\date{\today}

\section{Introduction}
For a long time dissipation has mainly been considered as a nuisance which destroys the coherence of a quantum state. However, in the last decade dissipation has been turned into a tool and the dissipative attractor dynamics has been used to stabilize complex quantum states \cite{MuellerZoller2012}. Examples of such realizations are the stabilization of a Tonks-Girardeau gas of molecules \cite{SyassenDuerr2008} or the creation of entanglement in an ion chain \cite{BarreiroBlatt2011}.

With a proper engineering of the coupling to the reservoir, it is possible to drive the system towards a desired quantum state. Here we concentrate on setups which exhibit local particle losses. Such a setup is realized using cold atomic gases subjected to an electron beam \cite{GerickeOtt2007} or more recently using near-resonant optical tweezers \cite{CormanEsslinger2019}. In the setup with the electron beam the quantum Zeno effect has been realized \cite{BarmettlerKollath2011,BarontiniOtt2013} and more recently, the phenomenon of coherent perfect absorption has been observed \cite{MuellersOtt2018}.  
Theoretically, this setup has been studied extensively for weakly interacting bosons \cite{BrazhnyiOtt2009,ShchesnovichKonotop2010,ZezyulinOtt2012} and for the Bose-Hubbard model \cite{BarmettlerKollath2011,ShchesnovichMogilevtsev2010,WitthautWimberger2011,KieferEmmanouilidisSirker2017} with respect to different properties including transport phenomena.

More recently, interacting fermions with a single-site dissipative defect causing particle losses were investigated theoretically \cite{FroemlDiehl2019, FroemlDiehl2019b,DamanetDaley2019} and experimentally \cite{CormanEsslinger2019, LebratEsslinger2019}. It was found that the quantized transport of fermions survives in the presence of dissipative quantum dots \cite{CormanEsslinger2019, LebratEsslinger2019}. Additionally, the interplay between dissipation strength, coherence and interaction leads to the emergence of the fluctuation induced quantum Zeno effect \cite{FroemlDiehl2019, FroemlDiehl2019b}, signaled by a reduced rate of the total particle loss. In particular, the particle loss at the Fermi momentum is totally suppressed for repulsive interactions and a metastable state arises, which cannot be described by a thermal state. As different approximations have been employed in the aforementioned work, the question remains whether in a full treatment of both, the interaction and the dissipation, this metastable non-thermal state survives.  

In this article we address this question and establish the existence of a quasi-stationary regime in time by using numerically exact matrix product state (MPS) methods to simulate the dynamics of a system of spinless fermions with nearest-neighbour interaction exposed to local loss processes. We simulate the full non-equilibrium dynamics of the system with a variety of different interaction and dissipation strengths. In the following we first introduce the model in Sec.~\ref{sec:model} and explain the method we use to investigate this model in Sec.~\ref{sec:method}. In Sec.~\ref{sec:te} we present the time evolution of the system; we discuss the initial dynamics of the local density in Sec.~\ref{sec:ie}, the quantum Zeno effect in Sec.~\ref{sec:zeno} and the properties of the metastable solution in Sec.~\ref{sec:meta}. 

\section{Model}\label{sec:model}

\begin{figure}
\includegraphics[width=\linewidth]{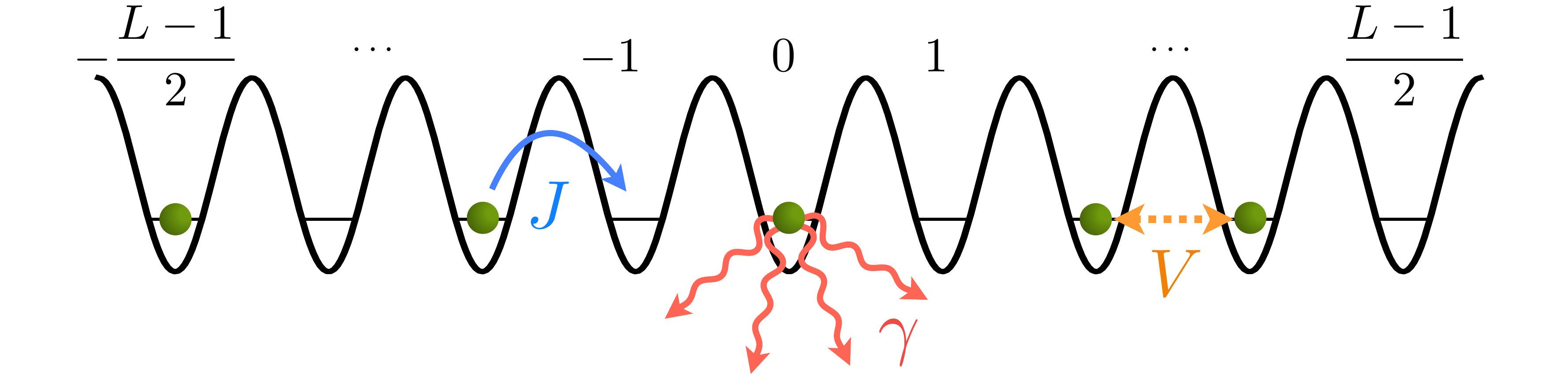}
\caption{Sketch of the set-up of interacting spinless fermions in a one-dimensional lattice that includes hopping between adjacent sites with amplitude $J$, nearest-neighbour interaction with strength $V$ and Markovian particle loss at the central site labeled as '0' with dissipation strength $\gamma$.} 
\label{fig:setup}
\end{figure}

We consider spinless fermions on a one dimensional lattice subjected to a local loss as sketched in Fig.~\ref{fig:setup}. Fermions on neighbouring sites are interacting with each other. The Hamiltonian describing the system is given by,
\begin{eqnarray}\label{eq:Ham}
H=&-&J \sum_{l=-\frac{L-1}{2}}^{\frac{L-1}{2}-1} \left (c^\dagger_{l} c^{\phantom{\dagger}}_{l+1}+\mathrm{H.c.} \right ) + V \sum_{l=-\frac{L-1}{2}}^{\frac{L-1}{2}-1}n_{l} n_{l+1}\nonumber\\
    &+&H_{\text{b}}.
\end{eqnarray}
For simplicity, we consider a chain of $L$ sites, where $L$ is an odd number. The operators $c_l^{\phantom{\dagger}}$ and $c_l^\dagger$ are the annihilation and creation operators for fermions at site $l$ and $n_l=c^\dagger_{l} c^{\phantom{\dagger}}_{l}$ is the local density operator for fermions at site $l$. The hopping amplitude and the interaction strength are denoted by $J$ and $V$, respectively. In order to decrease the boundary effects induced by the interaction the term
\begin{eqnarray}\label{eq:Ham_b}
H_{\text{b}}=&&V \left(n_{-\frac{L-1}{2}} + n_{\frac{L-1}{2}}\right) \frac{N}{L}
\end{eqnarray}
is added to the Hamiltonian. This term couples the density at the boundaries to the average value of the density $N/L$, where $N$ is the number of fermions, and therefore smoothens the boundary effects. For intermediate interaction $-2\le V/J \le 2$, the ground state of this model can be well described by a gapless Tomonaga-Luttinger liquid. At larger values of the interaction $ |V/J|> 2$, the systems enters a gapped phase and at repulsive interactions, charge density wave correlations are dominant \cite{Giamarchibook}.    

The chain of fermions is subjected at its center to a local particle loss. In ultracold atomic gases such losses can be created by the application of an electron beam \cite{BarontiniOtt2013} or a local radio frequency flip \cite{PalzerKoehl2009}. 
The system dynamics is described by a Lindblad master equation given by
\begin{align}\label{eq:Lindblad}
&\dot \rho(t) = -\frac{i}{\hbar} \left[ H, \rho(t) \right] + \gamma \left( c_0 \rho(t) c_0^\dagger - \frac{1}{2}\left\{\rho(t), c_0^\dagger c_0 \right\}  \right),
\end{align}
where $\rho$ is the density matrix of the fermionic system. Lost particles do not reenter the system. The first term on the right hand side describes the Hamiltonian evolution of the system with the Hamiltonian given above. The second term introduces the dissipative loss process with the so-called  jump operator given by  $c_0$  at the central site ($l=0$) with an amplitude $\gamma$.

\begin{figure*}
  \includegraphics[width=0.99\linewidth]{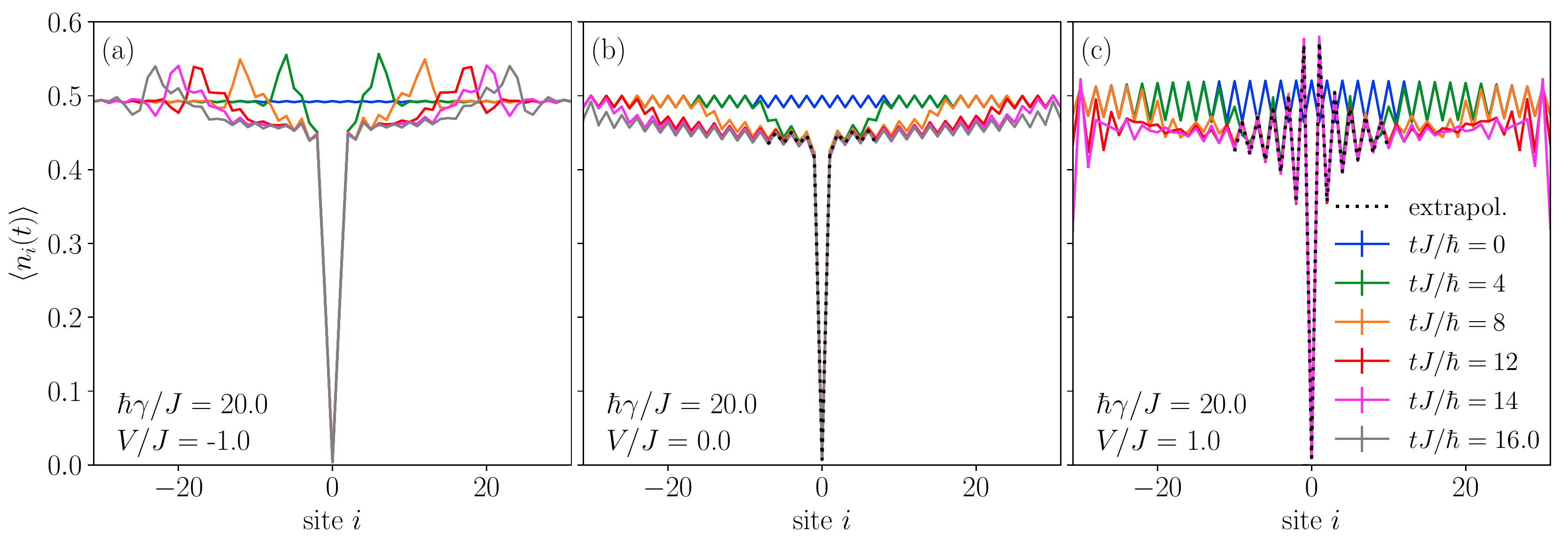}
\caption{\label{fig:ntime}Time evolution of the density profile starting from the ground state of $H$ with initial filling $N/L=(L-1)/(2L)$ after switching on the losses at $t=0$ for attractive, vanishing and repulsive interactions. The dotted line represents the extrapolated long-time metastable profiles, which emerge around the central site before the evolution is affected by boundary effects. Data is shown for a system of size $L=63$, different interaction strengths $V/J$, dissipation strength $\hbar \gamma/J=20$ and $10^3$ sampled trajectories.} 
\label{fig:density_evolution}
\end{figure*}

\section{Method}\label{sec:method}
We numerically simulate the dynamics of this open quantum system starting from the ground state of the Hamiltonian $H$, which is obtained by using the density matrix renormalization group (DMRG) method in the formulation of matrix product states (MPS) \cite{Schollwoeck2011}. The dissipative nature of the system is taken into account using the Monte-Carlo wave function method \cite{DalibardMolmer1992,  GardinerZoller1992, CarmichaelBook, BreuerPetruccione2002, Daley2014}. In this method the evolution of wave functions is calculated, rather than the evolution of the density matrix, at the cost of a stochastic sampling of many different trajectories. A single trajectory sample is created by a piecewise deterministic process in which a deterministic evolution, generated by a non-Hermitian Hamiltonian, here given by 
\begin{equation}
H_{\mathrm{eff}} = H - i \frac{\gamma}{2} c_0^\dagger c_0^{\phantom{\dagger}},
\end{equation}
is stochastically interrupted by applications of the jump operator. The duration of the deterministic evolution until the next jump occurs is the so-called waiting time $\tau$. This time is sampled according to the cumulative distribution 
\begin{equation}
P\left(\ket{\psi(t)}, \tau\right) = 1 - \left\lVert \exp\left( -i H_{\mathrm{eff}} \tau \right)\ket{\psi(t)}\right\rVert,
\end{equation}
specified by the decay of the norm of the evolving state, which originates from the non-unitary evolution \cite{BreuerPetruccione2002}.

The time-dependent expectation value of an observable is averaged over a sufficiently large number of stochastically sampled wave functions to get a desired accuracy. The evolution of the observable calculated by Monte-Carlo wave function method coincides with the results calculated by the evolution of the density matrix with Lindblad master equation (Eq.~\ref{eq:Lindblad}).
In our work, the deterministic part of the evolution is computed using time-dependent matrix product state (tMPS) methods, which efficiently approximate evolving quantum states while keeping a very high level of accuracy\cite{DaleyVidal2004, WhiteFeiguin2004}. This method relies on a truncation scheme, which is controlled by the convergence parameters of the truncation weight $\varepsilon$ and the bond dimension $D$ \cite{Schollwoeck2011, PaeckleHubig2019}. In the presented results we define a truncation goal $\varepsilon=10^{-12}$ that determines a maximum accepted truncation error, but never exceeds a defined maximum value for the bond dimension $D=300$. Additionally, within this method we approximate the time evolution operator of the effective Hamiltonian by a second order Trotter-Suzuki decomposition, which is controlled by the time step $\Delta t$ \cite{DaleyVidal2004, WhiteFeiguin2004}, chosen here as $\Delta t=0.05\hbar/J$ if not stated otherwise. We assure convergence of our results with respect to these parameters. 

In this work, the stochastic sampling is preferred over the full density matrix evolution which can be performed within MPS methods by the so-called purification \cite{Schollwoeck2011} for three reasons. Firstly, the initial state being the ground state of an interacting problem is a potentially highly entangled state. Thus, it  requires a very large bond dimension when formulated as a purified matrix product state making an accurate description very challenging. Secondly, the fact that the Markovian dissipation is only represented by a single jump operator makes the additional stochastic selection among many different operators at the jump times unnecessary. This will lead to smaller deviations between the trajectory samples. Thus the number of trajectories needed for the convergence of the observables is less compared to situations with many jump operators. Furthermore, due to the presence of the atom loss, the total number of particles is not conserved by the full Lindblad evolution of the density matrix, and all particle number sectors need to be taken into account. Contrary to that, in the stochastic simulation performed here, each part of the deterministic time-evolution with non-unitary Hamiltonian (Eq.~\ref{eq:Ham_b}) can be computed using the atom number as a conservation law to reduce the computational complexity. Only the applications of the  jump operator change the number of atoms.

\section{Time evolution of the interacting system subjected to losses}\label{sec:te}
The local loss will lead to an emptying of the system around the central site. In a finite system, the long time steady state is the trivial empty system. However, we will discuss the initial and intermediate time dynamics, at which an interesting metastable state arises in the system.

\subsection{Initial evolution of the density distribution}\label{sec:ie}
\begin{figure}
  \includegraphics[width=0.99\linewidth]{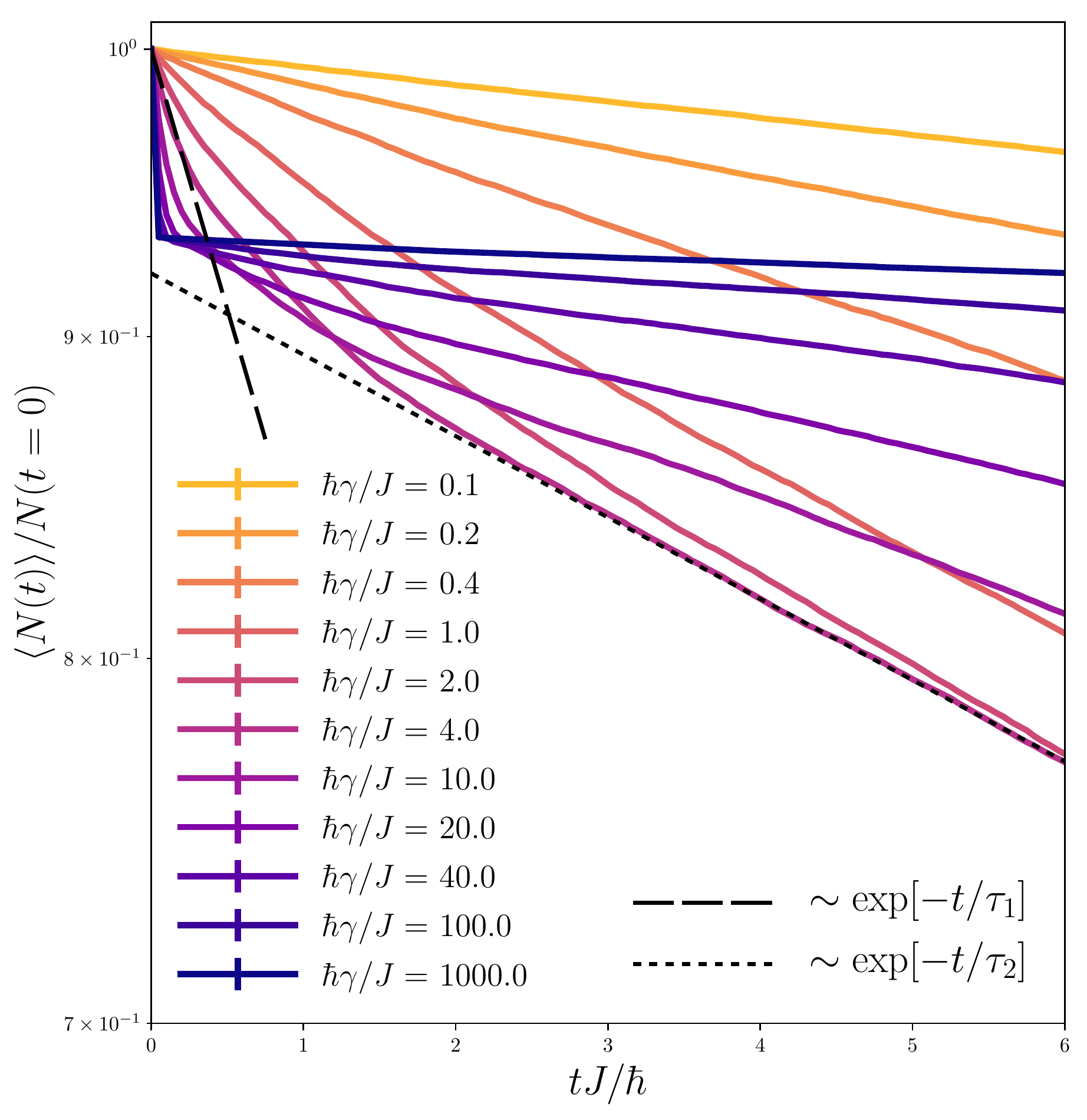}
\caption{\label{fig:Ntottime} Dependence of the evolution of the total particle number $N$ on the dissipation strength $\gamma$ for a system of size $L=15$, $N(t=0)=7$ and repulsive interaction strength $V/J=-1$ exhibiting two different time regimes. Dotted and dashed lines show exponential fits in the two time regions, where the exponents $1/\tau_{1,2}$ are associated with inverse time scales. We present an average over $10^4$ trajectories.} 
\label{fig:Ntot_dissipationdependence}
\end{figure}

\begin{figure}
  \includegraphics[width=0.99\linewidth]{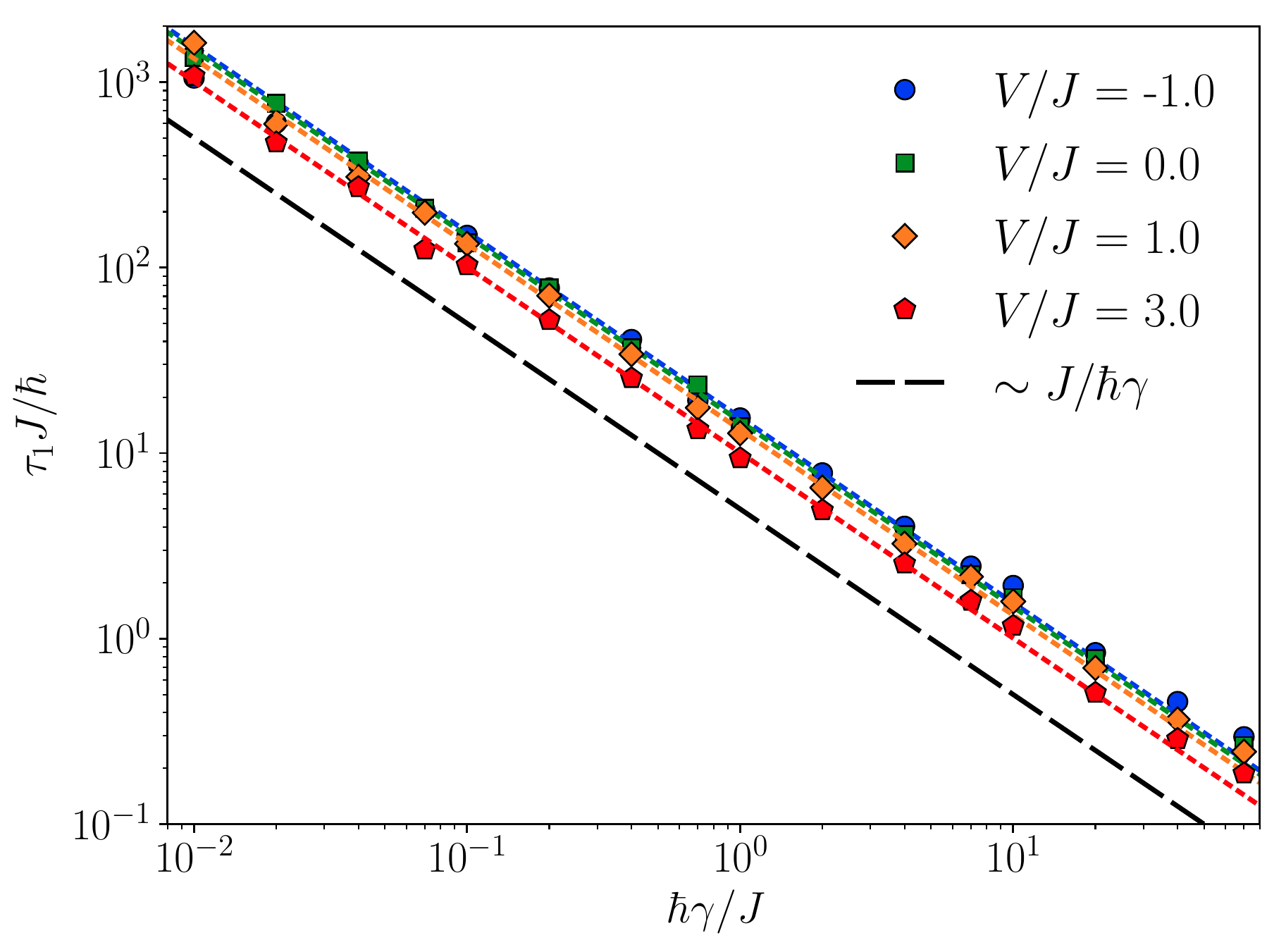}
\caption{\label{fig:tau1}  Inverse dependence of the time scale $\tau_1$ determining the decay of $N(t)$ in the first time regime on the dissipation strength $\gamma$ for different interaction strengths. The time scales were extracted by exponential fits as shown in Fig.~\ref{fig:Ntot_dissipationdependence}. Dotted lines are linear fits in $1/\gamma$ and the dashed line is a guide to the eye. We considered $L=15$ sites, an initial particle number $N(t=0)=7$ and $10^4$ trajectories. The time step was chosen as $\Delta t J/\hbar=0.01$ for $\hbar\gamma/J>20$ and $\Delta t J/\hbar=0.05$ else.} 
\end{figure}

In Fig.~\ref{fig:ntime} we show the evolution of the spatial density distribution of the fermions starting from the ground state of the Hamiltonian with $(L-1)/2$ particles. At time $t=0$ the local losses are switched on. Fig.~\ref{fig:ntime} shows that already in the initial density profile, oscillations are present which are induced by the open boundaries. These oscillations are stronger at repulsive interactions and are smeared out for attractive ones. After the switch-on of the losses, the density on the central site decreases first very quickly  as $n_0(t)\approx n_0(0)e^{-\alpha \gamma t}$, where $\alpha$ depends on the filling and the interaction strength, and then saturates to a metastable value at large dissipation. The saturation value $n_0(t\to\infty)$ depends on the dissipation and interaction strength. Here for the considered case of very strong dissipation $\gamma=20J/\hbar$ the saturation value of the density at central site  is very low. This initial process of the depletion of the central site by the loss can also be seen in the total number of atoms lost from the system as shown in Fig.~\ref{fig:Ntottime}. For very short times, the decay is exponential, i.e.~ $N(t)=N(0)e^{- t/\tau_1}$ with the time-scale depending strongly on $\gamma$. As shown in Fig.~\ref{fig:tau1}, the initial time-scale $\tau_1$ has linear dependence on the inverse of the dissipation strength, i.e.~ $\tau_1\propto 1/\gamma$. The slope depends on the interaction strength and the filling. For weak dissipation strength this regime lasts for very long times and defines the main dynamics of the system on the time scale accessible to the simulation. In contrast, for strong dissipation strength this regime lasts for very short times and crosses over to a second regime. 

\begin{figure}
\includegraphics[width=0.99\linewidth]{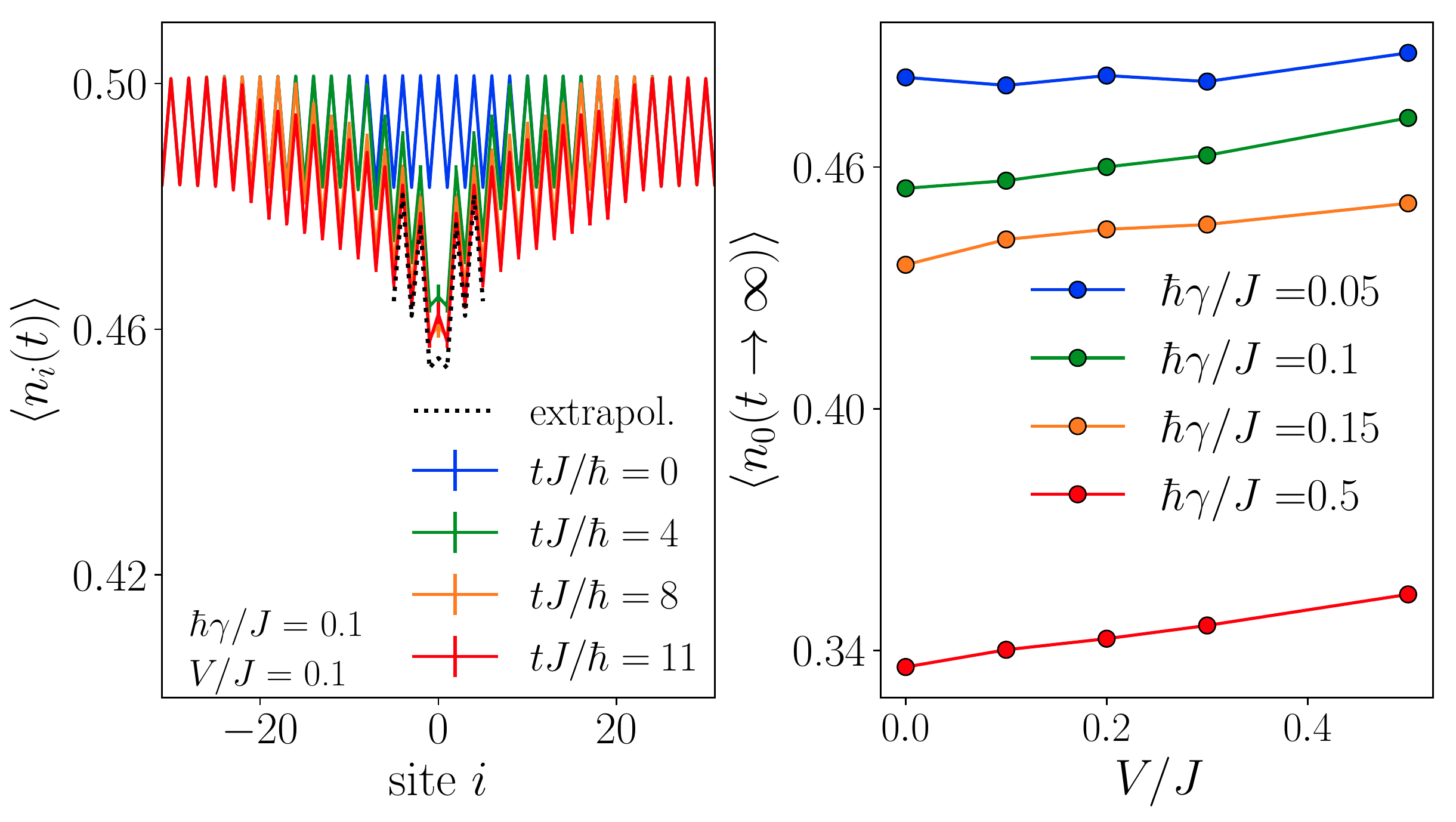}
\caption{Left panel: Time evolution of the density profile for comparably low interaction $V/J =0.1$ and low dissipation strength $\hbar\gamma/J=0.1$. The dashed line marks a $t\rightarrow\infty$ extrapolation extracted from an exponential fit with offset in the second time region. Right panel: Dependence of the metastable quasi-stationary state occupation of the central site on both dissipation and interaction strength represented by value at the longest time reachable by the simulation. A system of size $L=63$ has been considered that initially contains $N(t=0)=31$ particles using $2\cdot10^3$ trajectories.} 
\label{fig:n0}
\end{figure}

\subsection{Spreading of the depletion and the Zeno effect}\label{sec:zeno}
\begin{figure}
  \includegraphics[width=0.99\linewidth]{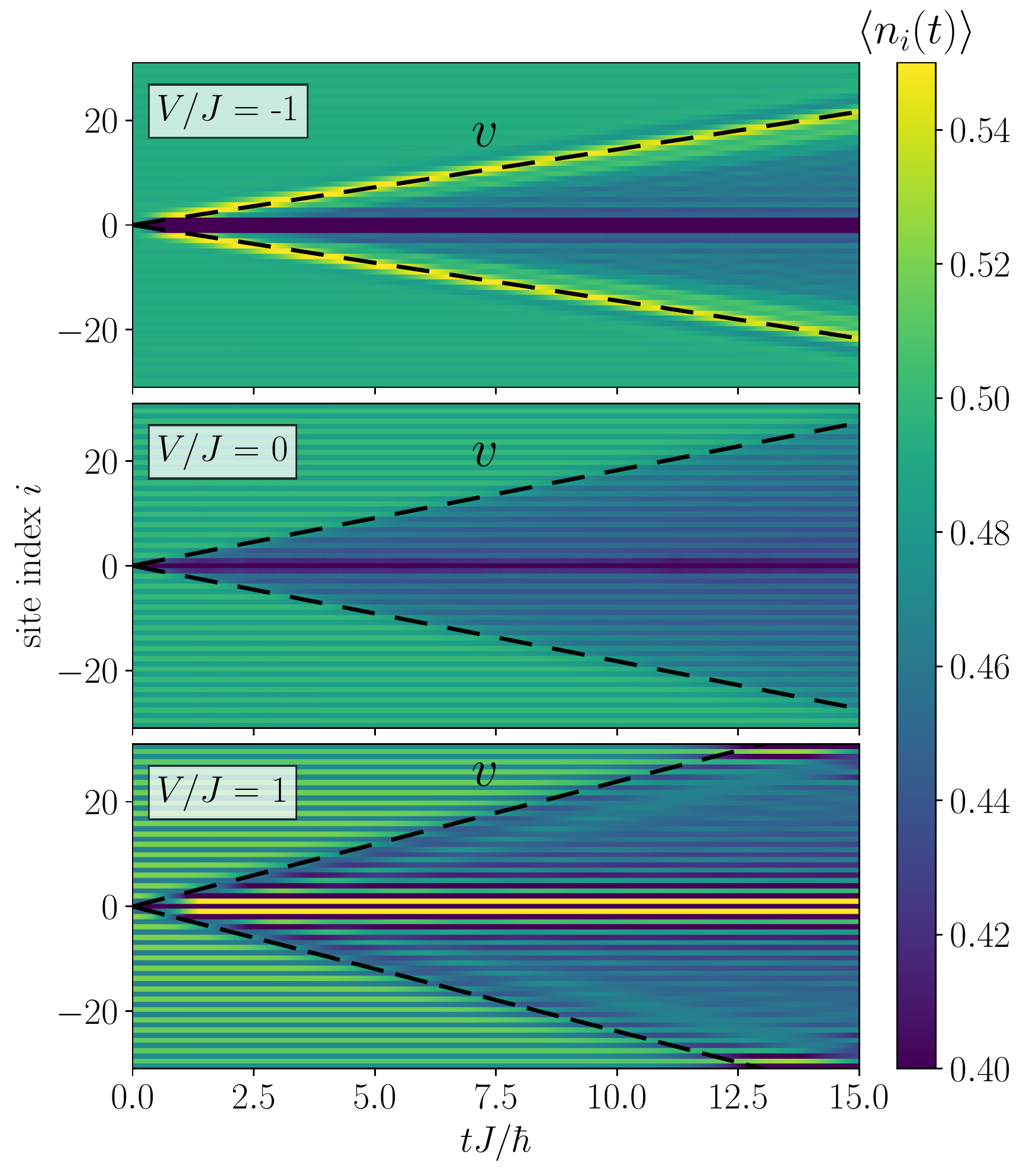}
\caption{Dissipative evolution of the density distribution for the parameters of Fig.~\ref{fig:density_evolution} showing a light cone structure bounded by an interaction dependent velocity $v$, indicated by the dashed line as a guide to the eye.} 
\label{fig:lightcone}
\end{figure}

After the initial depletion of the central site, the tunneling processes between the different sites set in and lead to a reduction of the density in a region around the central site (Fig.~\ref{fig:ntime}). For strong dissipation, this is signaled by a much slower decay of the particle number (Fig.~\ref{fig:Ntottime}). The region which is depleted spreads linearly with time through the system and a particle current towards the central site is induced (Fig.~\ref{fig:lightcone}). The velocity $v$ of the spreading depends strongly on the interaction strength and the filling. A slower spreading is found for the attractively interacting system compared to the non-interacting system and a faster spreading for the repulsive case. A metastable state is formed within the cone of the spreading. 

In the non-interacting case the spreading of the metastable state takes place by a reduction of the density. In contrast, the front of the spreading has very different nature for attractive or repulsive interactions. For the attractive case, a small density dip is followed by a high density peak which is propagating through the system. Behind this peak, the density is reduced and small density oscillations arise. For the repulsive interaction, the initially stronger density oscillations are perturbed, the density is reduced and in the metastable regime, strong density oscillations arise close to the lossy site.

\begin{figure}\label{fig:zeno}
\includegraphics[width=0.99\linewidth]{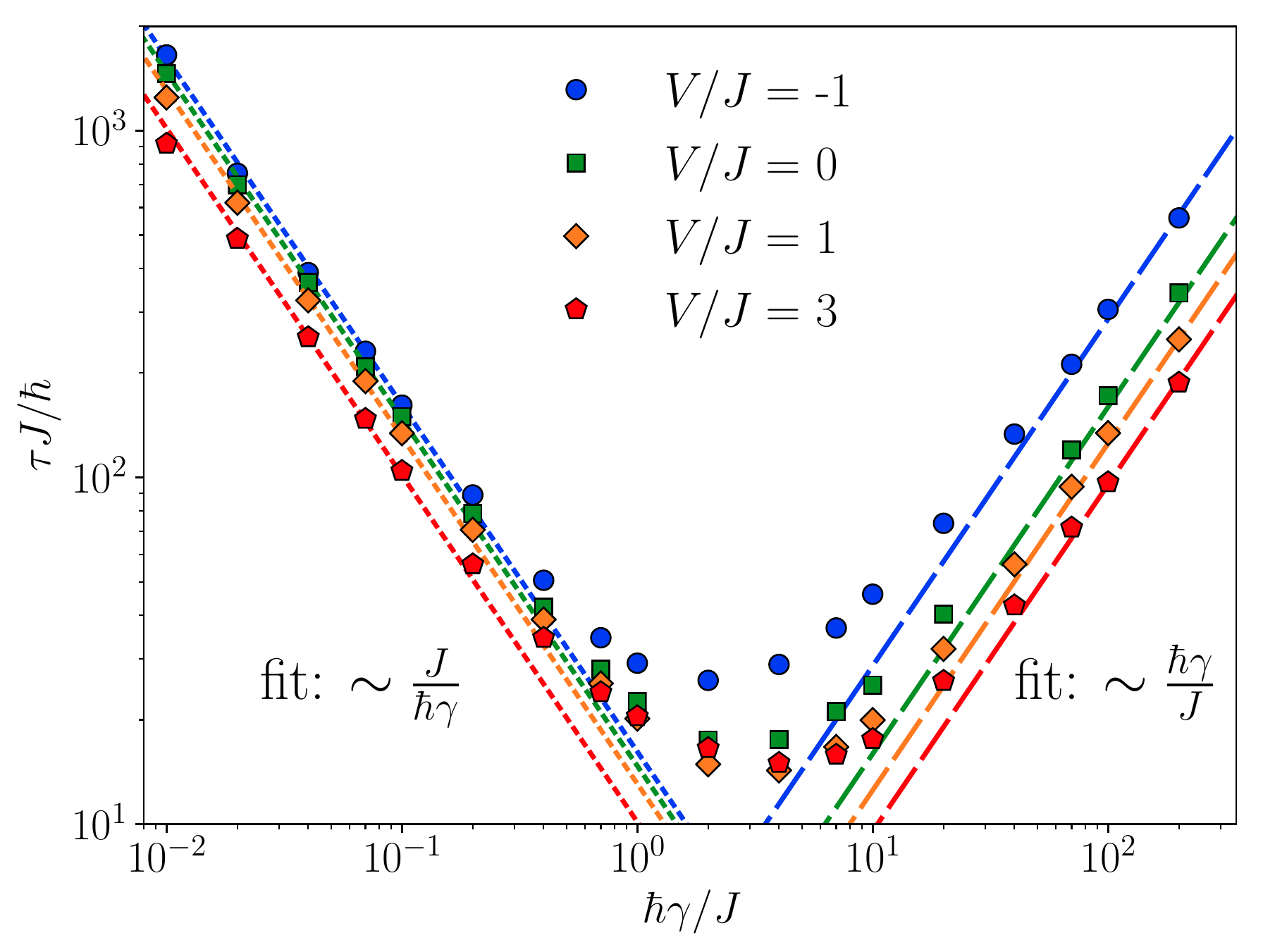}
\caption{Dependence of the time scale of the particle loss for intermediate times on the strength of the dissipative coupling $\gamma$ for different interaction strengths extracted from exponential fits of $\mathrm{e}^{-t/\tau}$ as shown in Fig.~\ref{fig:Ntot_dissipationdependence}. 
Dotted (dashed) lines represent fits for an inverse (a linear) dependence of the time scale on the dissipation strength $\gamma$ for the corresponding interaction. The linear decrease of the time scale for large values of $\gamma$ is the key signature for the quantum Zeno effect. System and simulation parameters are the same as in Fig.~\ref{fig:Ntot_dissipationdependence}.} 
\label{fig:zeno}
\end{figure}

The loss rate in the time-regime of the spreading is decreased for strong interaction and depends both on the dissipation and the interaction strength. In order to analyze this in more detail, we show in Fig.~\ref{fig:zeno} the time scale $\tau$ in this intermediate time regime before the light cone reaches the boundaries and finite size effects set in. 

In Fig.~\ref{fig:zeno} the dependence of the dimensionless time scale $\tau J/\hbar$ on the dissipation $\hbar\gamma/J$ is plotted for different interaction strengths. In the small dissipation regime $\gamma \lesssim J/\hbar$ this time scale decreases with increasing $\gamma$ as $1/\gamma$. This is naively expected, as the total loss rate is mainly set by the amplitude of the dissipative coupling. In this regime, the tunneling dynamics is faster than the time-scale of the loss. Only a relatively weak dependence on the interaction strength exists and can be addressed to the different velocities of the spreading of the depletion of the density through the system. At larger repulsive interaction a slower time-scale is found. 

In the strongly dissipative regime, the quantum Zeno effect occurs for all considered interaction strengths. The time scale of the loss increases again for higher dissipation proportional to $\gamma$. This increase of the time scale can be interpreted in the way that many measurements -- which correspond to the action of the dissipation -- freeze the quantum state of the system such that its evolution in time is prevented.
The occurrence of the quantum Zeno effect was previously predicted in an analogous setup with bosonic atoms \cite{BarmettlerKollath2011} and has been observed in Ref.~\cite{BarontiniOtt2013}. Our findings confirm also the findings for the non-interacting fermionic case and a different analytical approximation \cite{FroemlDiehl2019, FroemlDiehl2019b}. 
We further see a strong dependence of the decay time scale on the interaction strength. Although all interactions exhibit the quantum Zeno effect, the decay time scale depends on the interaction strength. The attractive fermions decay slower compared to non-interacting fermions and the repulsive fermions decay faster.


\subsection{Properties of the metastable state}\label{sec:meta}
As we have seen at intermediate times, a long lived metastable state is reached in the surrounding of the lossy site. Some properties of this metastable state have been analyzed previously using different methods in Ref.~\cite{FroemlDiehl2019,FroemlDiehl2019b}.  
This metastable state is very special in the sense that is does not resemble any ground or finite temperature state in equilibrium even considering the presence of a conservative potential at the site of the loss. This can already be seen from the real space density distribution. In the non-interacting case Fig.~\ref{fig:ntime}(b), a depletion of the density can be seen. The quasi-stationary value of the central site (Fig.~\ref{fig:n0}) shows a strong dependence on both the interaction and the dissipation strengths. Our results are roughly approximated by a linear behavior with the interaction strength for the considered parameter regime. A strong dissipative coupling leads to an almost complete emptying of the central site in the metastable state. Also a depletion around the central site becomes stable with the surprising occurrence of oscillations in space whose period does not vary with time. This is in contrast to the equilibrium properties of such a system. In equilibrium Friedel oscillations of the density are caused by an impurity which are of the form $n(x)\propto n_0\frac{\sin(2k_F x)}{x}$ where $n_{0}$ is the average density of the state and $k_F=\pi n_0$ is the Fermi wave vector. The oscillation period depends via the Fermi momentum linearly  on the inverse of the filling $n_0$. Thus, a depletion in equilibrium would lead to an increased oscillation frequency. However, in the metastable state such a change of the oscillation frequency is not observed. 

\begin{figure}
  \includegraphics[width=0.99\linewidth]{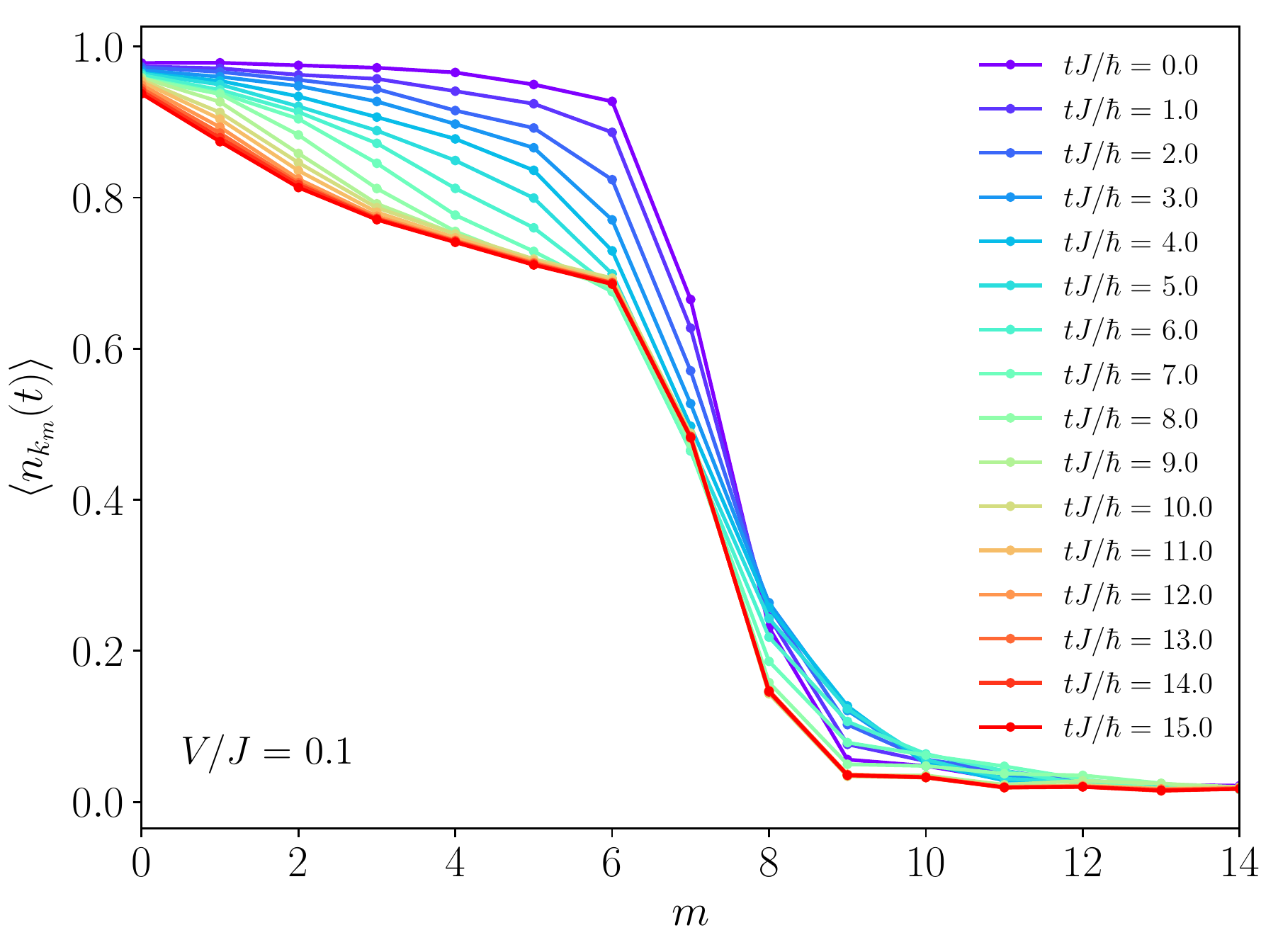}
   \includegraphics[width=0.99\linewidth]{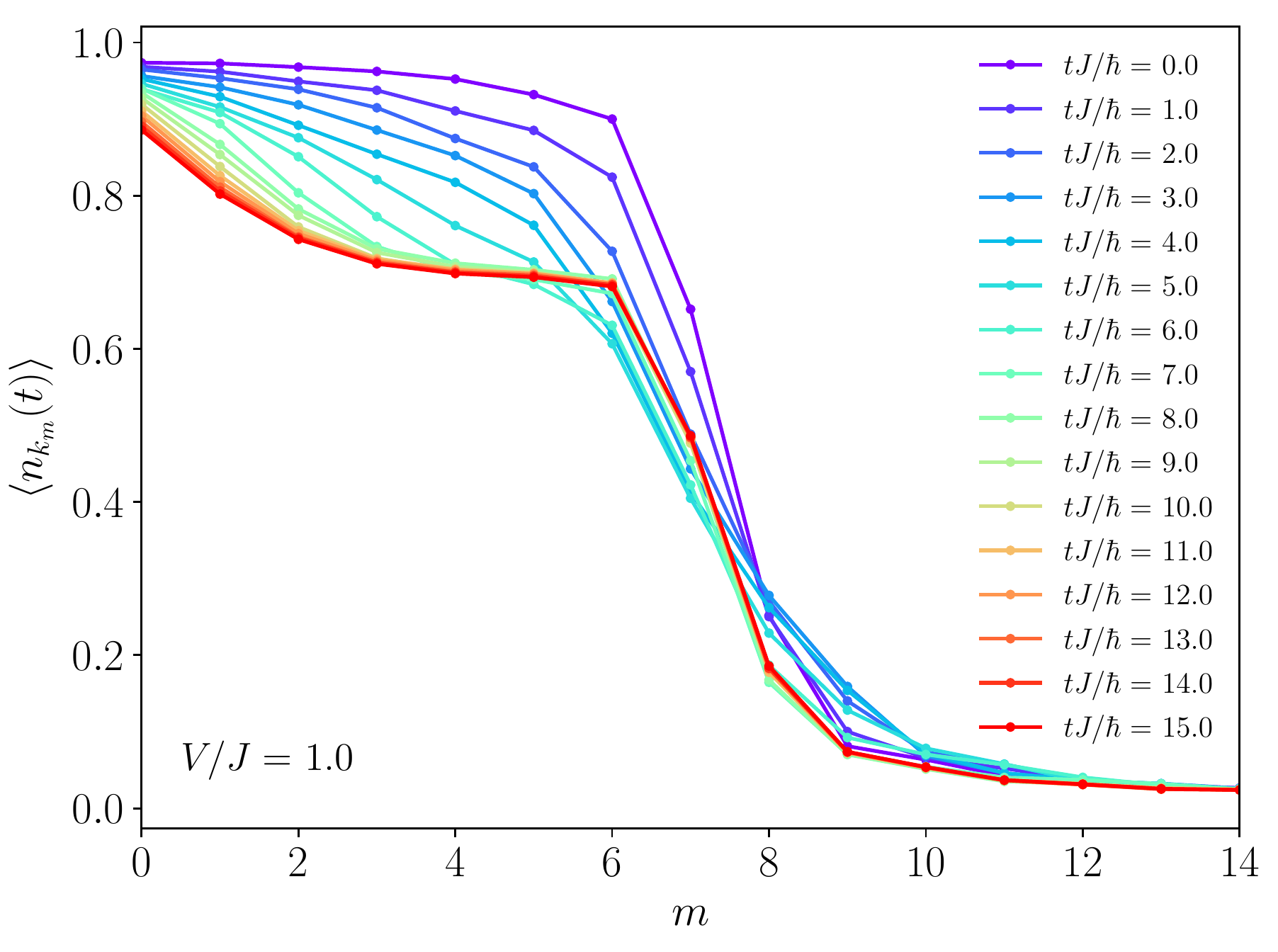}
\caption{Evolution of the momentum distribution represented by the Fourier transform of the density considering only $\tilde{L}=16$ sites on the right side of the defect in the center for $V/J=0.1$ (upper panel) and $V/J=1$ (lower panel). Results are presented for system size $L=63$, initial particle number $N(t=0)=31$, dissipation strength $\hbar \gamma/J=20$ and $2\cdot 10^3$ trajectory samples.} 
\label{fig:nk}
\end{figure}
Let us turn to the evolution of the momentum distribution, which further underlines the non-equilibrium nature of this state. The prediction from the non-interacting system and from an approximate treatment of the interacting system is that the discontinuity at the Fermi momentum $k_F$ remains intact and a depletion of the lower momenta takes place \cite{FroemlDiehl2019,FroemlDiehl2019b}. This is another clear non-equilibrium signature of the metastable state, since at equilibrium  $k_F$ behaves linearly with the filling. This is rationalized by the fact that the loss is local in real space, and therefore empties out all occupied momentum modes roughly homogeneously. For repulsive interactions the loss of particles at the Fermi momentum is suppressed due to the scattering off the Friedel oscillations. This effect was called the fluctuation induced quantum Zeno effect, since it is induced by the interaction. However, an important open question is whether the scattering induced by the interaction, once fully taken into account, will lead to a redistribution of the fermions towards a thermal state. 

In Fig.~\ref{fig:nk} we show the momentum distribution $n_k$ corresponding only to sites from $[0,\tilde L - 1]$, where $\tilde L$ marks  the extend of the metastable region right from the lossy site at the longest considered times. The momentum distribution is defined by $n_{k_m}=c^\dagger_{k_m} c^{\phantom{\dagger}}_{k_m}$, where $c_{k_m}$ denotes the Fourier transform of the fermionic operators
\begin{align}
&c_{k_m} = \frac{1}{\sqrt{\tilde L}} \sum_{j=1}^{\tilde L} 	c_j \mathrm{e}^{-i  k_m j}, \\\notag
 &\text{with } k_m = \frac{m\pi}{\tilde L+1}\text{ and } m \in \{1, 2, \ldots, \tilde L \}.
\end{align}

The distribution at time $t=0$ shows the Fermi step at the initial Fermi wave vector within the resolution of the finite system size. For small interaction strength of $V=0.1J$ and dissipation $\hbar \gamma= 20J$ fermions are lost at all momenta below the Fermi level. However, after a certain time a stable distribution arises for momenta around the initial Fermi momentum. In particular in this distribution,  the Fermi step remains pinned at its initial value, whereas the occupation below the Fermi momentum is reduced. Additionally, a substructure just below the Fermi momentum sets in, which causes a small rise of the occupation around these values of the momentum. 
For stronger interactions $V=J$, a similar evolution is seen. The stronger interaction can be recognized in a smearing out of  the step in the initial momentum distribution. With time, the occupation of all momenta below the Fermi level is reduced and a rise is seen for the occupation just above the Fermi momentum. At intermediate times, the evolution reaches a long lived metastable state for momenta close to the initial Fermi surface. In this quasi-stationary occupation, the Fermi step is still located around its initial value. A slight increase of the occupation just below the Fermi level arises, which is more pronounced than in the case of weak interaction. Therefore, the metastable state keeps its non-equilibrium nature for intermediate times.

The slight rise close to the Fermi momentum could have its origin in the fluctuation induced quantum Zeno effect \cite{FroemlDiehl2019, FroemlDiehl2019b}. A more direct verification of the fluctuation induced quantum Zeno effect could be gained by determining transport properties of the metastable state, which goes beyond the scope of the current work.

\section{conclusion}\label{sec:conc}
We have investigated interacting spinless fermions subjected to a local loss by numerically exact calculations using MPS methods.  Our numerically exact treatment for moderate system sizes complements the previous treatments relying on approximate methods for interacting fermions in the thermodynamic limit \cite{FroemlDiehl2019,FroemlDiehl2019b}. Initially an exponentially fast reduction of the density on the lossy site is found. In a second time regime the Zeno effect sets in for large dissipation strengths and causes a slow down of the losses. As typical for the Zeno regime, the time-scale for this second regime is inversely proportional to the dissipation strength and depends on the interaction strength. A metastable state is formed in a broad region around the impurity, which has no thermal counterpart. In particular, Friedel oscillations occur still at the period related to the initial Fermi momentum and the Fermi step in the momentum distribution remains at its initial position even though the density is strongly depleted. A slight increase of the momentum distribution close to the initial Fermi momentum is found, which hints at the existence of the fluctuation induced quantum Zeno effect expected for larger systems \cite{FroemlDiehl2019, FroemlDiehl2019b}. However, further investigations of transport properties are required to strengthen this support, and to identify the fluctuation induced quantum Zeno effect.

\section{Acknowledgments}
We thank H. Fr\"oml, A. Chiocchetta, and H.~Ott for fruitful discussions. We acknowledge funding from the Deutsche Forschungsgemeinschaft (DFG, German Research Foundation) under project number 277625399 - TRR 185 project B3 and project number 277146847 - project C04,C05, under the DFG Collaborative Research Center (CRC) 1238 and under Germany's Excellence Strategy – Cluster of Excellence Matter and Light for Quantum Computing (ML4Q) EXC 2004/1 – 390534769 and the European Research Council (ERC) under the Horizon 2020 research and innovation programme, grant agreement No. 648166 (Phonton) and grant agreement No. 647434 (DOQS).

\bibliography{swolff_local_loss}

\begin{thebibliography}{29}%
\makeatletter
\providecommand \@ifxundefined [1]{%
 \@ifx{#1\undefined}
}%
\providecommand \@ifnum [1]{%
 \ifnum #1\expandafter \@firstoftwo
 \else \expandafter \@secondoftwo
 \fi
}%
\providecommand \@ifx [1]{%
 \ifx #1\expandafter \@firstoftwo
 \else \expandafter \@secondoftwo
 \fi
}%
\providecommand \natexlab [1]{#1}%
\providecommand \enquote  [1]{``#1''}%
\providecommand \bibnamefont  [1]{#1}%
\providecommand \bibfnamefont [1]{#1}%
\providecommand \citenamefont [1]{#1}%
\providecommand \href@noop [0]{\@secondoftwo}%
\providecommand \href [0]{\begingroup \@sanitize@url \@href}%
\providecommand \@href[1]{\@@startlink{#1}\@@href}%
\providecommand \@@href[1]{\endgroup#1\@@endlink}%
\providecommand \@sanitize@url [0]{\catcode `\\12\catcode `\$12\catcode
  `\&12\catcode `\#12\catcode `\^12\catcode `\_12\catcode `\%12\relax}%
\providecommand \@@startlink[1]{}%
\providecommand \@@endlink[0]{}%
\providecommand \url  [0]{\begingroup\@sanitize@url \@url }%
\providecommand \@url [1]{\endgroup\@href {#1}{\urlprefix }}%
\providecommand \urlprefix  [0]{URL }%
\providecommand \Eprint [0]{\href }%
\providecommand \doibase [0]{http://dx.doi.org/}%
\providecommand \selectlanguage [0]{\@gobble}%
\providecommand \bibinfo  [0]{\@secondoftwo}%
\providecommand \bibfield  [0]{\@secondoftwo}%
\providecommand \translation [1]{[#1]}%
\providecommand \BibitemOpen [0]{}%
\providecommand \bibitemStop [0]{}%
\providecommand \bibitemNoStop [0]{.\EOS\space}%
\providecommand \EOS [0]{\spacefactor3000\relax}%
\providecommand \BibitemShut  [1]{\csname bibitem#1\endcsname}%
\let\auto@bib@innerbib\@empty
\bibitem [{\citenamefont {M\"uller}\ \emph {et~al.}(2012)\citenamefont
  {M\"uller}, \citenamefont {Diehl}, \citenamefont {Pupillo},\ and\
  \citenamefont {Zoller}}]{MuellerZoller2012}%
  \BibitemOpen
  \bibfield  {author} {\bibinfo {author} {\bibfnamefont {M.}~\bibnamefont
  {M\"uller}}, \bibinfo {author} {\bibfnamefont {S.}~\bibnamefont {Diehl}},
  \bibinfo {author} {\bibfnamefont {G.}~\bibnamefont {Pupillo}}, \ and\
  \bibinfo {author} {\bibfnamefont {P.}~\bibnamefont {Zoller}},\ }\bibfield
  {booktitle}  {\ \bibinfo {series}
  {Advances In Atomic, Molecular, and Optical Physics},\ \textbf {\bibinfo
  {volume} {61}},\ \bibinfo {pages} {1 } (\bibinfo {year} {2012})}\BibitemShut
  {NoStop}%
\bibitem [{\citenamefont {Syassen}\ \emph {et~al.}(2008)\citenamefont
  {Syassen}, \citenamefont {Bauer}, \citenamefont {Lettner}, \citenamefont
  {Volz}, \citenamefont {Dietze}, \citenamefont {Garc{\'\i}a-Ripoll},
  \citenamefont {Cirac}, \citenamefont {Rempe},\ and\ \citenamefont
  {D{\"u}rr}}]{SyassenDuerr2008}%
  \BibitemOpen
  \bibfield  {author} {\bibinfo {author} {\bibfnamefont {N.}~\bibnamefont
  {Syassen}}, \bibinfo {author} {\bibfnamefont {D.~M.}\ \bibnamefont {Bauer}},
  \bibinfo {author} {\bibfnamefont {M.}~\bibnamefont {Lettner}}, \bibinfo
  {author} {\bibfnamefont {T.}~\bibnamefont {Volz}}, \bibinfo {author}
  {\bibfnamefont {D.}~\bibnamefont {Dietze}}, \bibinfo {author} {\bibfnamefont
  {J.~J.}\ \bibnamefont {Garc{\'\i}a-Ripoll}}, \bibinfo {author} {\bibfnamefont
  {J.~I.}\ \bibnamefont {Cirac}}, \bibinfo {author} {\bibfnamefont
  {G.}~\bibnamefont {Rempe}}, \ and\ \bibinfo {author} {\bibfnamefont
  {S.}~\bibnamefont {D{\"u}rr}},\ }\href {\doibase 10.1126/science.1155309}
  {\bibfield  {journal} {\bibinfo  {journal} {Science}\ }\textbf {\bibinfo
  {volume} {320}},\ \bibinfo {pages} {1329} (\bibinfo {year} {2008})}\  \BibitemShut
  {NoStop}%
\bibitem [{\citenamefont {Barreiro}\ \emph {et~al.}(2011)\citenamefont
  {Barreiro}, \citenamefont {M\"uller}, \citenamefont {Schindler},
  \citenamefont {Nigg}, \citenamefont {Monz}, \citenamefont {Chwalla},
  \citenamefont {Hennrich}, \citenamefont {Roos}, \citenamefont {Zoller},\ and\
  \citenamefont {Blatt}}]{BarreiroBlatt2011}%
  \BibitemOpen
  \bibfield  {author} {\bibinfo {author} {\bibfnamefont {J.~T.}\ \bibnamefont
  {Barreiro}}, \bibinfo {author} {\bibfnamefont {M.}~\bibnamefont {M\"uller}},
  \bibinfo {author} {\bibfnamefont {P.}~\bibnamefont {Schindler}}, \bibinfo
  {author} {\bibfnamefont {D.}~\bibnamefont {Nigg}}, \bibinfo {author}
  {\bibfnamefont {T.}~\bibnamefont {Monz}}, \bibinfo {author} {\bibfnamefont
  {M.}~\bibnamefont {Chwalla}}, \bibinfo {author} {\bibfnamefont
  {M.}~\bibnamefont {Hennrich}}, \bibinfo {author} {\bibfnamefont {C.~F.}\
  \bibnamefont {Roos}}, \bibinfo {author} {\bibfnamefont {P.}~\bibnamefont
  {Zoller}}, \ and\ \bibinfo {author} {\bibfnamefont {R.}~\bibnamefont
  {Blatt}},\ }\href@noop {} {\bibfield  {journal} {\bibinfo  {journal}
  {Nature}\ }\textbf {\bibinfo {volume} {470}},\ \bibinfo {pages} {486}
  (\bibinfo {year} {2011})}\BibitemShut {NoStop}%
\bibitem [{\citenamefont {Gericke}\ \emph {et~al.}(2007)\citenamefont
  {Gericke}, \citenamefont {W{\"u}rtz}, \citenamefont {Reitz}, \citenamefont
  {Utfeld},\ and\ \citenamefont {Ott}}]{GerickeOtt2007}%
  \BibitemOpen
  \bibfield  {author} {\bibinfo {author} {\bibfnamefont {T.}~\bibnamefont
  {Gericke}}, \bibinfo {author} {\bibfnamefont {P.}~\bibnamefont {W{\"u}rtz}},
  \bibinfo {author} {\bibfnamefont {D.}~\bibnamefont {Reitz}}, \bibinfo
  {author} {\bibfnamefont {C.}~\bibnamefont {Utfeld}}, \ and\ \bibinfo {author}
  {\bibfnamefont {H.}~\bibnamefont {Ott}},\ }\href {\doibase
  10.1007/s00340-007-2862-9} {\bibfield  {journal} {\bibinfo  {journal}
  {Applied Physics B}\ }\textbf {\bibinfo {volume} {89}},\ \bibinfo {pages}
  {447} (\bibinfo {year} {2007})}\BibitemShut {NoStop}%
\bibitem [{\citenamefont {Corman}\ \emph {et~al.}(2019)\citenamefont {Corman},
  \citenamefont {Fabritius}, \citenamefont {H\"ausler}, \citenamefont {Mohan},
  \citenamefont {Dogra}, \citenamefont {Husmann}, \citenamefont {Lebrat}, and\
  \citenamefont {Esslinger}}]{CormanEsslinger2019}%
  \BibitemOpen
  \bibfield  {author} {\bibinfo {author} {\bibfnamefont {L.}~\bibnamefont
  {Corman}}, \bibinfo {author} {\bibfnamefont {P.}~\bibnamefont {Fabritius}},
  \bibinfo {author} {\bibfnamefont {S.}~\bibnamefont {H\"ausler}}, \bibinfo
  {author} {\bibfnamefont {J.}~\bibnamefont {Mohan}}, \bibinfo {author}
  {\bibfnamefont {L.~H.}\ \bibnamefont {Dogra}}, \bibinfo {author}
  {\bibfnamefont {D.}~\bibnamefont {Husmann}}, \bibinfo {author} {\bibfnamefont
  {M.}~\bibnamefont {Lebrat}}, \ and\ \bibinfo {author} {\bibfnamefont
  {T.}~\bibnamefont {Esslinger}},\ }\href@noop {}
  \Eprint {http://arxiv.org/abs/1907.06436}
  {arXiv:1907.06436}\ (\bibinfo {year} {2019}) \BibitemShut {NoStop}%
\bibitem [{\citenamefont {Barmettler}\ and\ \citenamefont
  {Kollath}(2011)}]{BarmettlerKollath2011}%
  \BibitemOpen
  \bibfield  {author} {\bibinfo {author} {\bibfnamefont {P.}~\bibnamefont
  {Barmettler}}\ and\ \bibinfo {author} {\bibfnamefont {C.}~\bibnamefont
  {Kollath}},\ }\href {http://link.aps.org/doi/10.1103/PhysRevA.84.041606}
  {\bibfield  {journal} {\bibinfo  {journal} {Phys. Rev. A}\ }\textbf {\bibinfo
  {volume} {84}},\ \bibinfo {pages} {041606} (\bibinfo {year}
  {2011})}\BibitemShut {NoStop}%
\bibitem [{\citenamefont {Barontini}\ \emph {et~al.}(2013)\citenamefont
  {Barontini}, \citenamefont {Labouvie}, \citenamefont {Stubenrauch},
  \citenamefont {Vogler}, \citenamefont {Guarrera},\ and\ \citenamefont
  {Ott}}]{BarontiniOtt2013}%
  \BibitemOpen
  \bibfield  {author} {\bibinfo {author} {\bibfnamefont {G.}~\bibnamefont
  {Barontini}}, \bibinfo {author} {\bibfnamefont {R.}~\bibnamefont {Labouvie}},
  \bibinfo {author} {\bibfnamefont {F.}~\bibnamefont {Stubenrauch}}, \bibinfo
  {author} {\bibfnamefont {A.}~\bibnamefont {Vogler}}, \bibinfo {author}
  {\bibfnamefont {V.}~\bibnamefont {Guarrera}}, \ and\ \bibinfo {author}
  {\bibfnamefont {H.}~\bibnamefont {Ott}},\ }\href {\doibase
  10.1103/PhysRevLett.110.035302} {\bibfield  {journal} {\bibinfo  {journal}
  {Phys. Rev. Lett.}\ }\textbf {\bibinfo {volume} {110}},\ \bibinfo {pages}
  {035302} (\bibinfo {year} {2013})}\BibitemShut {NoStop}%
\bibitem [{\citenamefont {M{\"u}llers}\ \emph {et~al.}(2018)\citenamefont
  {M{\"u}llers}, \citenamefont {Santra}, \citenamefont {Baals}, \citenamefont
  {Jiang}, \citenamefont {Benary}, \citenamefont {Labouvie}, \citenamefont
  {Zezyulin}, \citenamefont {Konotop},\ and\ \citenamefont
  {Ott}}]{MuellersOtt2018}%
  \BibitemOpen
  \bibfield  {author} {\bibinfo {author} {\bibfnamefont {A.}~\bibnamefont
  {M{\"u}llers}}, \bibinfo {author} {\bibfnamefont {B.}~\bibnamefont {Santra}},
  \bibinfo {author} {\bibfnamefont {C.}~\bibnamefont {Baals}}, \bibinfo
  {author} {\bibfnamefont {J.}~\bibnamefont {Jiang}}, \bibinfo {author}
  {\bibfnamefont {J.}~\bibnamefont {Benary}}, \bibinfo {author} {\bibfnamefont
  {R.}~\bibnamefont {Labouvie}}, \bibinfo {author} {\bibfnamefont {D.~A.}\
  \bibnamefont {Zezyulin}}, \bibinfo {author} {\bibfnamefont {V.~V.}\
  \bibnamefont {Konotop}}, \ and\ \bibinfo {author} {\bibfnamefont
  {H.}~\bibnamefont {Ott}},\ }\href {\doibase 10.1126/sciadv.aat6539}
  {\bibfield  {journal} {\bibinfo  {journal} {Science Advances}\ }\textbf
  {\bibinfo {volume} {4}} (\bibinfo {year} {2018})}\BibitemShut {NoStop}%
\bibitem [{\citenamefont {Brazhnyi}\ \emph {et~al.}(2009)\citenamefont
  {Brazhnyi}, \citenamefont {Konotop}, \citenamefont {P\'erez-Garc\'\i{}a},\
  and\ \citenamefont {Ott}}]{BrazhnyiOtt2009}%
  \BibitemOpen
  \bibfield  {author} {\bibinfo {author} {\bibfnamefont {V.~A.}\ \bibnamefont
  {Brazhnyi}}, \bibinfo {author} {\bibfnamefont {V.~V.}\ \bibnamefont
  {Konotop}}, \bibinfo {author} {\bibfnamefont {V.~M.}\ \bibnamefont
  {P\'erez-Garc\'\i{}a}}, \ and\ \bibinfo {author} {\bibfnamefont
  {H.}~\bibnamefont {Ott}},\ }\href@noop {} {\bibfield  {journal} {\bibinfo
  {journal} {Phys. Rev. Lett.}\ }\textbf {\bibinfo {volume} {102}},\ \bibinfo
  {pages} {144101} (\bibinfo {year} {2009})}\BibitemShut {NoStop}%
\bibitem [{\citenamefont {Shchesnovich}\ and\ \citenamefont
  {Konotop}(2010)}]{ShchesnovichKonotop2010}%
  \BibitemOpen
  \bibfield  {author} {\bibinfo {author} {\bibfnamefont {V.~S.}\ \bibnamefont
  {Shchesnovich}}\ and\ \bibinfo {author} {\bibfnamefont {V.~V.}\ \bibnamefont
  {Konotop}},\ }\href {http://link.aps.org/OPTdoi/10.1103/PhysRevA.81.053611}
  {\bibfield  {journal} {\bibinfo  {journal} {Phys. Rev. A}\ }\textbf {\bibinfo
  {volume} {81}},\ \bibinfo {pages} {053611} (\bibinfo {year}
  {2010})}\BibitemShut {NoStop}%
\bibitem [{\citenamefont {Zezyulin}\ \emph {et~al.}(2012)\citenamefont
  {Zezyulin}, \citenamefont {Konotop}, \citenamefont {Barontini},\ and\
  \citenamefont {Ott}}]{ZezyulinOtt2012}%
  \BibitemOpen
  \bibfield  {author} {\bibinfo {author} {\bibfnamefont {D.~A.}\ \bibnamefont
  {Zezyulin}}, \bibinfo {author} {\bibfnamefont {V.~V.}\ \bibnamefont
  {Konotop}}, \bibinfo {author} {\bibfnamefont {G.}~\bibnamefont {Barontini}},
  \ and\ \bibinfo {author} {\bibfnamefont {H.}~\bibnamefont {Ott}},\ }\href
  {\doibase 10.1103/PhysRevLett.109.020405} {\bibfield  {journal} {\bibinfo
  {journal} {Phys. Rev. Lett.}\ }\textbf {\bibinfo {volume} {109}},\ \bibinfo
  {pages} {020405} (\bibinfo {year} {2012})}\BibitemShut {NoStop}%
\bibitem [{\citenamefont {Shchesnovich}\ and\ \citenamefont
  {Mogilevtsev}(2010)}]{ShchesnovichMogilevtsev2010}%
  \BibitemOpen
  \bibfield  {author} {\bibinfo {author} {\bibfnamefont {V.~S.}\ \bibnamefont
  {Shchesnovich}}\ and\ \bibinfo {author} {\bibfnamefont {D.~S.}\ \bibnamefont
  {Mogilevtsev}},\ }\href
  {http://link.aps.org/OPTdoi/10.1103/PhysRevA.82.043621} {\bibfield  {journal}
  {\bibinfo  {journal} {Phys. Rev. A}\ }\textbf {\bibinfo {volume} {82}},\
  \bibinfo {pages} {043621} (\bibinfo {year} {2010})}\BibitemShut {NoStop}%
\bibitem [{\citenamefont {Witthaut}\ \emph {et~al.}(2011)\citenamefont
  {Witthaut}, \citenamefont {Trimborn}, \citenamefont {Hennig}, \citenamefont
  {Kordas}, \citenamefont {Geisel},\ and\ \citenamefont
  {Wimberger}}]{WitthautWimberger2011}%
  \BibitemOpen
  \bibfield  {author} {\bibinfo {author} {\bibfnamefont {D.}~\bibnamefont
  {Witthaut}}, \bibinfo {author} {\bibfnamefont {F.}~\bibnamefont {Trimborn}},
  \bibinfo {author} {\bibfnamefont {H.}~\bibnamefont {Hennig}}, \bibinfo
  {author} {\bibfnamefont {G.}~\bibnamefont {Kordas}}, \bibinfo {author}
  {\bibfnamefont {T.}~\bibnamefont {Geisel}}, \ and\ \bibinfo {author}
  {\bibfnamefont {S.}~\bibnamefont {Wimberger}},\ }\href
  {http://link.aps.org/OPTdoi/10.1103/PhysRevA.83.063608} {\bibfield  {journal}
  {\bibinfo  {journal} {Phys. Rev. A}\ }\textbf {\bibinfo {volume} {83}},\
  \bibinfo {pages} {063608} (\bibinfo {year} {2011})}\BibitemShut {NoStop}%
\bibitem [{\citenamefont {Kiefer-Emmanouilidis}\ and\ \citenamefont
  {Sirker}(2017)}]{KieferEmmanouilidisSirker2017}%
  \BibitemOpen
  \bibfield  {author} {\bibinfo {author} {\bibfnamefont {M.}~\bibnamefont
  {Kiefer-Emmanouilidis}}\ and\ \bibinfo {author} {\bibfnamefont
  {J.}~\bibnamefont {Sirker}},\ }\href {\doibase 10.1103/PhysRevA.96.063625}
  {\bibfield  {journal} {\bibinfo  {journal} {Phys. Rev. A}\ }\textbf {\bibinfo
  {volume} {96}},\ \bibinfo {pages} {063625} (\bibinfo {year}
  {2017})}\BibitemShut {NoStop}%
\bibitem [{\citenamefont {Fr\"oml}\ \emph
  {et~al.}(2019{\natexlab{a}})\citenamefont {Fr\"oml}, \citenamefont
  {Chiocchetta}, \citenamefont {Kollath},\ and\ \citenamefont
  {Diehl}}]{FroemlDiehl2019}%
  \BibitemOpen
  \bibfield  {author} {\bibinfo {author} {\bibfnamefont {H.}~\bibnamefont
  {Fr\"oml}}, \bibinfo {author} {\bibfnamefont {A.}~\bibnamefont
  {Chiocchetta}}, \bibinfo {author} {\bibfnamefont {C.}~\bibnamefont
  {Kollath}}, \ and\ \bibinfo {author} {\bibfnamefont {S.}~\bibnamefont
  {Diehl}},\ }\href {\doibase 10.1103/PhysRevLett.122.040402} {\bibfield
  {journal} {\bibinfo  {journal} {Phys. Rev. Lett.}\ }\textbf {\bibinfo
  {volume} {122}},\ \bibinfo {pages} {040402} (\bibinfo {year}
  {2019}{\natexlab{a}})}\BibitemShut {NoStop}%
\bibitem [{\citenamefont {Fr\"oml}\ \emph
  {et~al.}(2019{\natexlab{b}})\citenamefont {Fr\"oml}, \citenamefont {Muckel},
  \citenamefont {Kollath}, \citenamefont {Chiocchetta},\ and\ \citenamefont
  {Diehl}}]{FroemlDiehl2019b}%
  \BibitemOpen
  \bibfield  {author} {\bibinfo {author} {\bibfnamefont {H.}~\bibnamefont
  {Fr\"oml}}, \bibinfo {author} {\bibfnamefont {C.}~\bibnamefont {Muckel}},
  \bibinfo {author} {\bibfnamefont {C.}~\bibnamefont {Kollath}}, \bibinfo
  {author} {\bibfnamefont {A.}~\bibnamefont {Chiocchetta}}, \ and\ \bibinfo
  {author} {\bibfnamefont {S.}~\bibnamefont {Diehl}},\ } \Eprint
  {http://arxiv.org/abs/1910.10741} {arXiv:1910.10741}
  (\bibinfo {year} {2019}{\natexlab{b}})
  \BibitemShut {NoStop}%
\bibitem [{\citenamefont {Damanet}\ \emph {et~al.}(2019)\citenamefont
  {Damanet}, \citenamefont {Mascarenhas}, \citenamefont {Pekker},\ and\
  \citenamefont {Daley}}]{DamanetDaley2019}%
  \BibitemOpen
  \bibfield  {author} {\bibinfo {author} {\bibfnamefont {F.}~\bibnamefont
  {Damanet}}, \bibinfo {author} {\bibfnamefont {E.}~\bibnamefont
  {Mascarenhas}}, \bibinfo {author} {\bibfnamefont {D.}~\bibnamefont {Pekker}},
  \ and\ \bibinfo {author} {\bibfnamefont {A.~J.}\ \bibnamefont {Daley}},\
  },\ \Eprint
  {http://arxiv.org/abs/1904.03631} {arXiv:1904.03631}
  \href@noop {} (\bibinfo {year} {2019}) \BibitemShut
  {NoStop}%
\bibitem [{\citenamefont {Lebrat}\ \emph {et~al.}(2019)\citenamefont {Lebrat},
  \citenamefont {H\"ausler}, \citenamefont {Fabritius}, \citenamefont
  {Husmann}, \citenamefont {Corman},\ and\ \citenamefont
  {Esslinger}}]{LebratEsslinger2019}%
  \BibitemOpen
  \bibfield  {author} {\bibinfo {author} {\bibfnamefont {M.}~\bibnamefont
  {Lebrat}}, \bibinfo {author} {\bibfnamefont {S.}~\bibnamefont {H\"ausler}},
  \bibinfo {author} {\bibfnamefont {P.}~\bibnamefont {Fabritius}}, \bibinfo
  {author} {\bibfnamefont {D.}~\bibnamefont {Husmann}}, \bibinfo {author}
  {\bibfnamefont {L.}~\bibnamefont {Corman}}, \ and\ \bibinfo {author}
  {\bibfnamefont {T.}~\bibnamefont {Esslinger}},\ } \Eprint
  {http://arxiv.org/abs/1902.05516} {arXiv:1902.05516},\ \href@noop {} (\bibinfo {year} {2019})
  \BibitemShut {NoStop}%
\bibitem [{\citenamefont {Giamarchi}(2004)}]{Giamarchibook}%
  \BibitemOpen
  \bibfield  {author} {\bibinfo {author} {\bibfnamefont {T.}~\bibnamefont
  {Giamarchi}},\ }\href@noop {} {\emph {\bibinfo {title} {Quantum Physics in
  One Dimension}}}\ (\bibinfo  {publisher} {Oxford University Press},\ \bibinfo
  {address} {Oxford},\ \bibinfo {year} {2004})\BibitemShut {NoStop}%
\bibitem [{\citenamefont {Palzer}\ \emph {et~al.}(2009)\citenamefont {Palzer},
  \citenamefont {Zipkes}, \citenamefont {Sias},\ and\ \citenamefont
  {K\"ohl}}]{PalzerKoehl2009}%
  \BibitemOpen
  \bibfield  {author} {\bibinfo {author} {\bibfnamefont {S.}~\bibnamefont
  {Palzer}}, \bibinfo {author} {\bibfnamefont {C.}~\bibnamefont {Zipkes}},
  \bibinfo {author} {\bibfnamefont {C.}~\bibnamefont {Sias}}, \ and\ \bibinfo
  {author} {\bibfnamefont {M.}~\bibnamefont {K\"ohl}},\ }\href {\doibase
  10.1103/PhysRevLett.103.150601} {\bibfield  {journal} {\bibinfo  {journal}
  {Phys. Rev. Lett.}\ }\textbf {\bibinfo {volume} {103}},\ \bibinfo {pages}
  {150601} (\bibinfo {year} {2009})}\BibitemShut {NoStop}%
\bibitem [{\citenamefont {Schollw{\"o}ck}(2011)}]{Schollwoeck2011}%
  \BibitemOpen
  \bibfield  {author} {\bibinfo {author} {\bibfnamefont {U.}~\bibnamefont
  {Schollw{\"o}ck}},\ }\href {\doibase
  http://dx.doi.org/10.1016/j.aop.2010.09.012} {\bibfield  {journal} {\bibinfo
  {journal} {Annals of Physics}\ }\textbf {\bibinfo {volume} {326}},\ \bibinfo
  {pages} {96 } (\bibinfo {year} {2011})}\BibitemShut {NoStop}%
\bibitem [{\citenamefont {Dalibard}\ \emph {et~al.}(1992)\citenamefont
  {Dalibard}, \citenamefont {Castin},\ and\ \citenamefont
  {M\o{}lmer}}]{DalibardMolmer1992}%
  \BibitemOpen
  \bibfield  {author} {\bibinfo {author} {\bibfnamefont {J.}~\bibnamefont
  {Dalibard}}, \bibinfo {author} {\bibfnamefont {Y.}~\bibnamefont {Castin}}, \
  and\ \bibinfo {author} {\bibfnamefont {K.}~\bibnamefont {M\o{}lmer}},\
  }\href@noop {} {\bibfield  {journal} {\bibinfo  {journal} {Phys. Rev. Lett.}\
  }\textbf {\bibinfo {volume} {68}},\ \bibinfo {pages} {580} (\bibinfo {year}
  {1992})}\BibitemShut {NoStop}%
\bibitem [{\citenamefont {Gardiner}\ \emph {et~al.}(1992)\citenamefont
  {Gardiner}, \citenamefont {Parkins},\ and\ \citenamefont
  {Zoller}}]{GardinerZoller1992}%
  \BibitemOpen
  \bibfield  {author} {\bibinfo {author} {\bibfnamefont {C.~W.}\ \bibnamefont
  {Gardiner}}, \bibinfo {author} {\bibfnamefont {A.~S.}\ \bibnamefont
  {Parkins}}, \ and\ \bibinfo {author} {\bibfnamefont {P.}~\bibnamefont
  {Zoller}},\ }\href {\doibase 10.1103/PhysRevA.46.4363} {\bibfield  {journal}
  {\bibinfo  {journal} {Phys. Rev. A}\ }\textbf {\bibinfo {volume} {46}},\
  \bibinfo {pages} {4363} (\bibinfo {year} {1992})}\BibitemShut {NoStop}%
\bibitem [{\citenamefont {Carmichael}(1991)}]{CarmichaelBook}%
  \BibitemOpen
  \bibfield  {author} {\bibinfo {author} {\bibfnamefont {H.}~\bibnamefont
  {Carmichael}},\ }\href@noop {} {\emph {\bibinfo {title} {An open systems
  approach to quantum optics}}}\ (\bibinfo  {publisher} {Springer Verlag},\
  \bibinfo {address} {Berlin Heidelberg},\ \bibinfo {year} {1991})\BibitemShut
  {NoStop}%
\bibitem [{\citenamefont {Breuer}\ and\ \citenamefont
  {Petruccione}(2002)}]{BreuerPetruccione2002}%
  \BibitemOpen
  \bibfield  {author} {\bibinfo {author} {\bibfnamefont {H.~P.}\ \bibnamefont
  {Breuer}}\ and\ \bibinfo {author} {\bibfnamefont {F.}~\bibnamefont
  {Petruccione}},\ }\href@noop {} {\emph {\bibinfo {title} {The theory of open
  quantum systems}}}\ (\bibinfo  {publisher} {Oxford University Press},\
  \bibinfo {address} {Oxford},\ \bibinfo {year} {2002})\BibitemShut {NoStop}%
\bibitem [{\citenamefont {Daley}(2014)}]{Daley2014}%
  \BibitemOpen
  \bibfield  {author} {\bibinfo {author} {\bibfnamefont {A.~J.}\ \bibnamefont
  {Daley}},\ }\href {\doibase 10.1080/00018732.2014.933502} {\bibfield
  {journal} {\bibinfo  {journal} {Advances in Physics}\ }\textbf {\bibinfo
  {volume} {63}},\ \bibinfo {pages} {77} (\bibinfo {year} {2014})} \BibitemShut {NoStop}%
\bibitem [{\citenamefont {Daley}\ \emph {et~al.}(2004)\citenamefont {Daley},
  \citenamefont {Kollath}, \citenamefont {Schollw\"ock},\ and\ \citenamefont
  {Vidal}}]{DaleyVidal2004}%
  \BibitemOpen
  \bibfield  {author} {\bibinfo {author} {\bibfnamefont {A.~J.}\ \bibnamefont
  {Daley}}, \bibinfo {author} {\bibfnamefont {C.}~\bibnamefont {Kollath}},
  \bibinfo {author} {\bibfnamefont {U.}~\bibnamefont {Schollw\"ock}}, \ and\
  \bibinfo {author} {\bibfnamefont {G.}~\bibnamefont {Vidal}},\ }\href@noop {}
  {\bibfield  {journal} {\bibinfo  {journal} {J.~ Stat.~ Mech.: Theor.~ Exp.~}\
  }\textbf {\bibinfo {volume} {P04005}} (\bibinfo {year} {2004})}\BibitemShut
  {NoStop}%
\bibitem [{\citenamefont {White}\ and\ \citenamefont
  {Feiguin}(2004)}]{WhiteFeiguin2004}%
  \BibitemOpen
  \bibfield  {author} {\bibinfo {author} {\bibfnamefont {S.~R.}\ \bibnamefont
  {White}}\ and\ \bibinfo {author} {\bibfnamefont {A.~E.}\ \bibnamefont
  {Feiguin}},\ }\href@noop {} {\bibfield  {journal} {\bibinfo  {journal}
  {Phys.~ Rev.~ Lett.}\ }\textbf {\bibinfo {volume} {93}},\ \bibinfo {pages}
  {076401} (\bibinfo {year} {2004})}\BibitemShut {NoStop}%
\bibitem [{\citenamefont {Paeckel}\ \emph {et~al.}(2019)\citenamefont
  {Paeckel}, \citenamefont {K\"ohler}, \citenamefont {Swoboda}, \citenamefont
  {Manmana}, \citenamefont {Schollw\"ock},\ and\ \citenamefont
  {Hubig}}]{PaeckleHubig2019}%
  \BibitemOpen
  \bibfield  {author} {\bibinfo {author} {\bibfnamefont {S.}~\bibnamefont
  {Paeckel}}, \bibinfo {author} {\bibfnamefont {T.}~\bibnamefont {K\"ohler}},
  \bibinfo {author} {\bibfnamefont {A.}~\bibnamefont {Swoboda}}, \bibinfo
  {author} {\bibfnamefont {S.~R.}\ \bibnamefont {Manmana}}, \bibinfo {author}
  {\bibfnamefont {U.}~\bibnamefont {Schollw\"ock}}, \ and\ \bibinfo {author}
  {\bibfnamefont {C.}~\bibnamefont {Hubig}},\ }\Eprint {http://arxiv.org/abs/1901.05824} {arXiv:1901.05824}
  \href@noop {} {\  (\bibinfo{year} {2019})}
  \BibitemShut {NoStop}%
\end{thebibliography}%

\end{document}